\documentclass[aps,prl,showpacs,nofootinbib,amsmath,superscriptaddress,preprintnumbers,reprint]{revtex4-2}
\usepackage{graphicx}
\usepackage{dcolumn}
\usepackage{amsmath,amssymb}
\usepackage{bbold}
\usepackage{color}
\usepackage[dvipsnames]{xcolor}
\usepackage{amsmath}
\usepackage{ulem}
\definecolor{darkblue}{rgb}{0,0,0.6}
\usepackage[colorlinks,linkcolor=darkblue,citecolor=darkblue,urlcolor=darkblue]{hyperref}
\usepackage{soul}
\definecolor{mygreen}{rgb}{0.0,0.55,0.3}

\renewcommand{\epsilon}{\varepsilon}

\begin{document}

\title{Directed percolation transition to active turbulence driven by non-reciprocal forces}

\author{Juliane U. Klamser}

\affiliation{Laboratoire Charles Coulomb (L2C), Université de Montpellier \& CNRS (UMR 5221), 34095 Montpellier, France}

\author{Ludovic Berthier}

\affiliation{Gulliver, CNRS UMR 7083, ESPCI Paris, PSL Research University, 75005 Paris, France}

\date{\today}

\begin{abstract}
We numerically study the collective dynamics of dense particle assemblies driven by non-reciprocal pairwise forces of amplitude $\kappa$. At a critical value $\kappa_{\rm c}$, the system undergoes a dynamical phase transition from an absorbing state ($\kappa < \kappa_{\rm c}$) to a chaotic steady state ($\kappa > \kappa_{\rm c}$). The chaotic phase is marked by nontrivial spatiotemporal velocity correlations and mixing, reminiscent of active turbulence in self-propelled systems. The sharp onset of chaos shows critical scaling consistent with the universality class of directed percolation. We argue that this transition is generic to a broad class of locally-driven, dense disordered materials.
\end{abstract}

\maketitle

{\it Introduction--}The dynamics of dense disordered systems driven by non-conservative forces is a broad topic, covering from the non-equilibrium dynamics of spin glasses~\cite{CrisantiSompolinsky1987,GutfreundYoung1988,SpitznerKinzel1989} and neural networks~\cite{Hopfield1982,SompolinskySommers1988} to the rheology of soft glassy materials~\cite{sollich1997rheology}, with more recent studies from the field of active matter~\cite{berthier2019glassy}. The interplay between a complex energy landscape characterized by multiple metastable states and non-equilibrium driving forces makes the problem difficult and broadly relevant~\cite{CugliandoloPeliti1997}. Here, we ask how a glass state obtained in a two-dimensional dense assembly of repulsive particles responds to the presence of non-reciprocal pairwise forces.  

In the mean-field limit, the driven dynamics of disordered glassy models becomes analytically tractable and is well understood. This generic problem has been attacked from multiple angles, from mean-field models of brain dynamics~\cite{SompolinskySommers1988}, spin~\cite{CrisantiSompolinsky1987,lorenzana2024non}, structural~\cite{berthier2000two} and active~\cite{berthier2013non,morse2021direct} glasses, interfaces in the presence of quenched disorder~\cite{cugliandolo1996large}, and models for complex ecological systems~\cite{Bunin2017,RoyBunin2020,RosBunin2023,AltieriBiroli2021,ArnoulxDePireyBunin2024}. In this limit, the interplay between marginally-stable glassy states and driving forces typically leads to emergent chaotic dynamics even for arbitrary-low driving amplitude and low temperatures. Arrested glassy states and aging dynamics are thus generically suppressed by non-conservative forces~\cite{kurchan1997rheology}. In addition, the distinction between global (such as external fields and shear flow) and local (such as asymmetric local interactions) is immaterial, as space is unimportant in the mean-field limit. 

The interest for local non-reciprocal forces recently gained new momentum due to developments in the field of active matter where the role and influence of non-reciprocal interactions has led to a surge of research activity \cite{BowickRamaswamy2022}. This is primarily motivated by novel experimental realizations of many-body non-equilibrium systems driven by non-reciprocal forces~\cite{LisinHyde2020,GrauerLiebchen2021,BililignIrvine2022,GuptaRamaswamy2022,GuilletBartolo2025}. It was also discovered that local non-reciprocal driving forces may differ conceptually from global forcing and can lead to new emergent physical phenomena (odd transport coefficients, dynamical phase transitions)~\cite{FruchartVitelli2021,MarkovichLubensky2021,FurchartVitelli2023}. Different physical settings have been explored, including mixtures of particle species~\cite{SotoGolestanian2014,IvlevLoewen2015,Agudo-CanalejoGolestanian2019,NasouriGolestanian2020,saha2020scalar,you2020nonreciprocity,Ouazan-ReboulGolestanian2021,GrauerLiebchen2021,FruchartVitelli2021,Frohoff-HulsmannThiele2021,Frohoff-HulsmannThiele2021_2,ZhangGarcia-Millan2023,DinelliTailleur2023,MartinVitelli2024}, single-component bulk systems with short-range asymmetric couplings~\cite{BarberisPeruani2016,GrossmannLutz2014,SahaGolestanian2019,KnezevicStark2022,PoncetBartolo2022,GuilletBartolo2025,BililignIrvine2022,CohenGolestanian2014,GranekSolon2020,loos2023longrange,rouzaire2025nonreciprocal,popli2025ordering,dopierala2025inescapable}, active–passive mixtures where activity induces non-reciprocal effective interactions~\cite{BanerjeeRao2022,GuptaRamaswamy2022,GulatiMarchetti2024}, and long-ranged couplings mediated by active fluids~\cite{GranekSolon2020,BaekKafri2018}.

We address two main questions. In contrast to mean-field predictions of chaos at arbitrarily weak forcing, is there instead a sharp transition at finite forcing amplitude in finite-dimensional glassy particle systems? What is the nature of the chaotic steady state created by non-reciprocal forces in dense particle systems? To answer them, we construct a two-dimensional particle model where (reciprocal) repulsive interactions lead to the formation of amorphous glasses, which are then driven by local non-reciprocal pairwise forces. This setting allows us to directly study the competition between glassiness and non-reciprocal driving forces in finite dimensions. 

Our analysis first demonstrates the existence of a sharp phase transition at a critical value $\kappa_{\rm c}$ of non-reciprocal forces between an arrested dynamics below $\kappa_{\rm c}$, and a chaotic steady state above. We conclude that in finite dimensions, glassy states and aging survive a finite amount of local non-reciprocal forces, in sharp contrast to dynamical mean-field theory predictions. We demonstrate in addition that this non-equilibrium phase transition at $\kappa_{\rm c}$ is accompanied by diverging timescales and lengthscales, with a criticality consistent with the universality class of directed percolation~\cite{Hinrichsen2000}. Our second central result is that chaotic dynamics above $\kappa_{\rm c}$ is characterized by spatio-temporal velocity correlations and particle transport that qualitatively resemble observations in active turbulent systems~\cite{alert2022active}. The agreement becomes quantitative with models of self-propelled fluids~\cite{Mesoscale2024}. 

These results are relevant in the context of driven glassy states, as they establish the existence of a sharp phase transition driven by local driving, that resembles the depinning~\cite{Fisher1998,Chauve2000creep} and yielding transitions~\cite{Nicolas2018deformation,berthier2025yielding} observed when the drive is global. They are also relevant in the context of active turbulence, where sharp transitions to chaotic flows have been numerically observed in active nematics~\cite{DoostmohammadiYeomans2017,HillebrandAlert2025}, strengthening analogies with inertial turbulence where directed percolation transitions have also been identified~\cite{Pomeau1986,SiposGoldenfeld2011,LemoultHof2016,SanoTamai2016,Hof2023,ShihGoldenfeld2025}.    

\begin{figure}
\includegraphics{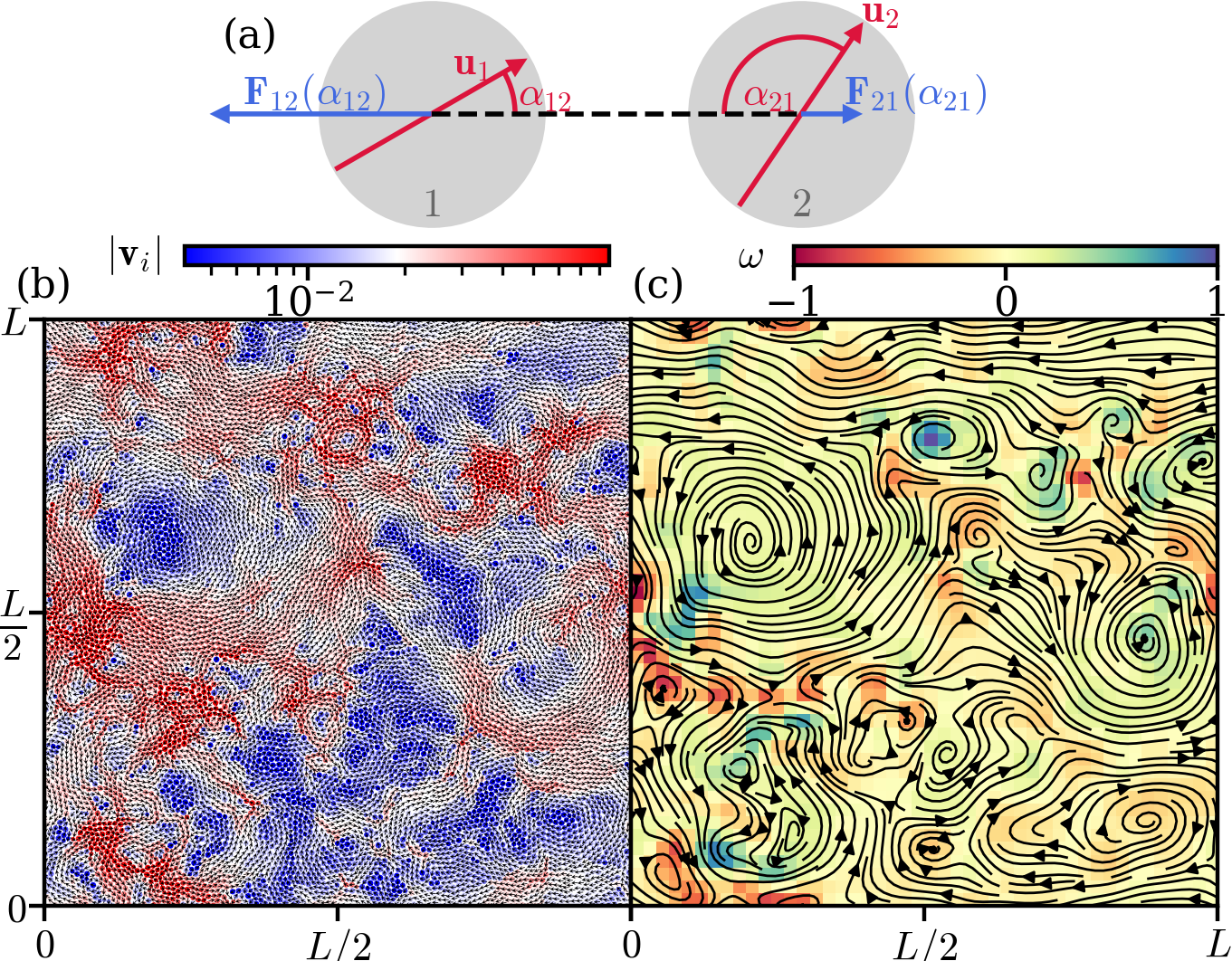}
\caption{(a) The repulsive forces $\mathbf{F}_{12}$ and $\mathbf{F}_{21}$ between particles 1 and 2 with orientations ${\bf u}_1$ and ${\bf u}_2$ are modulated by the angles $\alpha_{12} \neq \alpha_{21}$, resulting in non-reciprocal pairwise interactions. The orientations act as an `eye' for the particles. (b) Typical velocity field in chaotic steady state at large enough $\kappa=0.08 > \kappa_{\rm c}$ (colors code for amplitudes, arrows for orientations). Large scale correlated regions of fast and slow velocities are organized along streams and vortices. (c) Corresponding vorticity field, with stream lines.}
\label{fig:Model_VelocityField_DyeMix} 
\end{figure}

{\it Model--}We consider a model of $N$ interacting particles characterized by a position ${\bf r}_i$ and an orientation ${\bf u}_i$, evolving in a two-dimensional square box of linear size $L$ with periodic boundaries. The particle positions $\mathbf{r}_i$ follow an athermal overdamped dynamics
\begin{equation}
   \gamma \dot{\mathbf{r}}_i(t) = \sum_{j\ne i} \mathbf{F}_{ij}(r_{ij},\alpha_{ij})\,,
    \label{Eq:EqOfMotion}
\end{equation}
where $\gamma$ is the friction coefficient and $\mathbf{F}_{ij}(r_{ij},\alpha_{ij})$ is the force particle $j$ exerts on particle $i$, see Fig.~\ref{fig:Model_VelocityField_DyeMix}(a). The pairwise forces depend both on the interparticle distance $r_{ij} = |\mathbf{r}_i-\mathbf{r}_j| = |\mathbf{r}_{ij}|$ and on the angle 
\begin{equation}
\alpha_{ij} = \arccos\left(\frac{\mathbf{r}_{ji}\cdot\mathbf{u}_i}{|\mathbf{r}_{ji}|}\right) ,
\end{equation}
defined between $\mathbf{r}_{ji}$ and the orientation $\mathbf{u}_i = (\cos\theta_i,\sin\theta_i)$ of particle $i$. In general $\alpha_{ij}\ne\alpha_{ji}$, and this asymmetry is used to modulate the amplitude of the forces so that $|\mathbf{F}_{ij}|\ne|\mathbf{F}_{ji}|$, thus explicitly breaking the action-reaction principle at the level of each pairwise interaction. In practice we choose
\begin{equation}
    \mathbf{F}_{ij}(r_{ij}, \alpha_{ij}) =
    \left(1+\kappa \cos\alpha_{ij}\right)\, \mathbf{F}^{\rm rep}_{ij}(r_{ij})\,,
    \label{Eq:DefForce}
\end{equation}
where $\mathbf{F}^{\rm rep}_{ij}(r_{ij}) = - \boldsymbol{\nabla}_i U(r_{ij})$ is a repulsive conservative force. The non-dimensional parameter $\kappa$ tunes the amount of non-reciprocity, and the dynamics is conservative for $\kappa=0$. The repulsive potential is 
the standard Weeks-Chandler-Andersen potential $U(r_{ij})=4 \varepsilon\left[\left(\sigma_{i j} / r_{i j}\right)^{12}-\left(\sigma_{i j} / r_{i j}\right)^6+1 / 4\right]$ for $r_{ij} < 2^{1/6}\sigma_{ij}$ and $0$ otherwise, where $\varepsilon$ is an energy scale and $\sigma_{ij} = (\sigma_i + \sigma_j)/2$, with $\sigma_i$ the diameter of particle $i$. For $\kappa = 0$, particles interact via repulsive forces that do not depend on orientations. When $\kappa > 0$ particles experience a stronger repulsion from frontal neighbors ($\alpha_{ij} \simeq 0$) than from rear neighbors ($\alpha_{ij} \simeq \pi$). In other words, the direction $\theta_i$ is used to emulate the idea of a vision cone~\cite{CouzinNigel2002,BarberisPeruani2016,DurveSayeed2018}, often used in models of active matter. Equation~(\ref{Eq:DefForce}) should directly produce run and chase motion at the pair level. Since forces are short-ranged, the model is only interesting when the density is large enough for each particle to interact with several neighbors. 

To suppress translational order and focus on disordered fluid or glass states, we draw $\sigma_i$ from a uniform distribution over the interval  $[0.8\sigma, 1.2\sigma]$. The particle orientations are also uniformly distributed over $[0,2\pi]$. In practice, we draw the angles as $\theta_i = (i - 1) 2 \pi /N$ to minimize center of mass motion. Orientations are permanently frozen. Since Eq.~(\ref{Eq:EqOfMotion}) does not conserve the total momentum,  velocities are measured in the frame of the center of mass, $\mathbf{v}_i(t) = \dot{\mathbf{r}}_i(t)-1/N\sum \dot{\mathbf{r}}_i(t)$. We report results for the largest system size considered, $N = 12800$, using $L=140 \sigma$. Results are reported using $\sigma$ as the unit length and $\sigma^2 \gamma / \epsilon$ as the unit time.

{\it Emerging collective flows--}For $\kappa = 0$, the dynamics in Eq.~\eqref{Eq:EqOfMotion} reduces to gradient descent in a potential energy, thus driving the system towards one of the many local minima of the potential energy landscape. In such minimum, reciprocal forces balance each other and mechanical equilibrium is reached. The system forms a zero-temperature amorphous solid: a glass. When $\kappa > 0$, however, the dynamics (\ref{Eq:EqOfMotion}) is non-conservative and the system no longer minimizes a global energy function. The system can still find a force-balanced configuration, where the only remaining motion is the one of the center of mass. Such state does not minimize an energy function, but is still a frozen glass. Another possibility is that forces never balance, and the system remains in a chaotic steady state where particles constantly move and rearrange. The latter certainly exists at large enough $\kappa$, where the net force on each particle is, on average, opposite to $\mathbf{u}_i$, thus persistently pushing particles backwards through the elastic surrounding medium. This competition between local driving and glassiness, which we aimed to capture in our model, gives rise at large $\kappa$ to spatially correlated velocity fields, as illustrated in Fig.~\ref{fig:Model_VelocityField_DyeMix}(b) and further emphasized by the structured vorticity field shown in Fig.~\ref{fig:Model_VelocityField_DyeMix}(c). These extended spatial correlations are unrelated to the orientation field $\theta_i$ which remains fully disordered at all state points (not shown). 

\begin{figure}[t]
\includegraphics{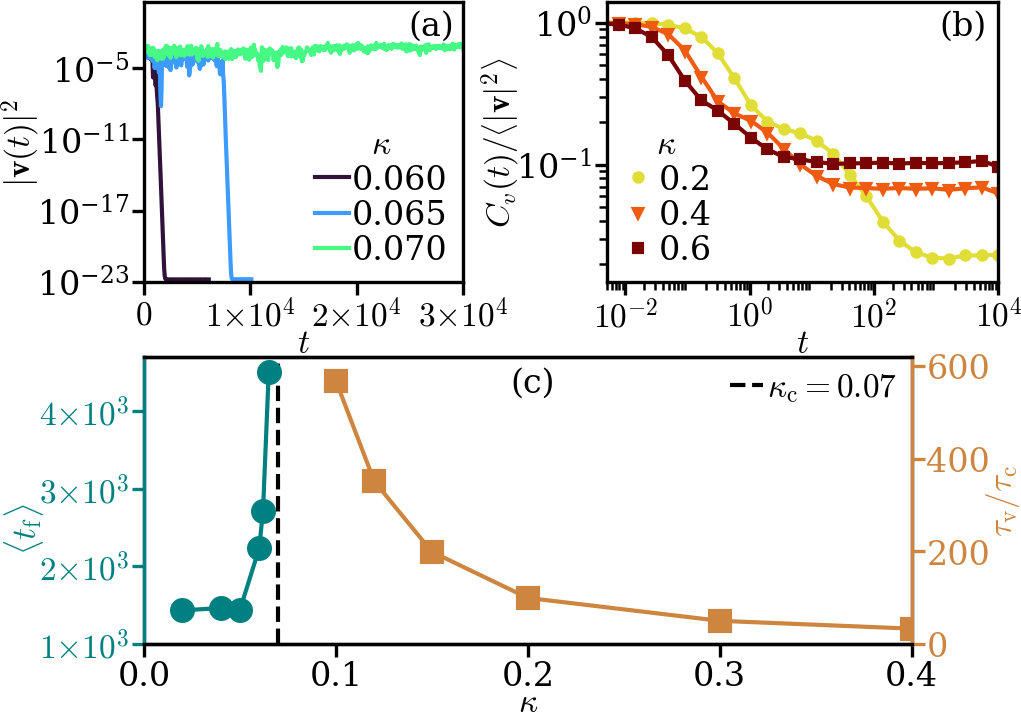}
\caption{(a) Time evolution of the mean-squared velocity for different values of $\kappa$, starting from random initial conditions. The time  $t_{\rm f}$ to enter an absorbing state with vanishing velocity increases with $\kappa$. Above $\kappa_{\rm c}$, the system remains active within the numerical observation time. (b) Normalized velocity autocorrelation for different $\kappa$, showing a two-step decay to a constant, defining a microscopic collision timescale $\tau_{\rm c}$, and a much longer persistent time $\tau_{\rm v}$.(c) The mean freezing time $\langle t_{\rm f} \rangle$ and the rescaled persistent time $\tau_{\rm v} / \tau_{\rm c}$ diverge on both sides of the absorbing phase transition at $\kappa_{\rm c}$.}
\label{fig:Freezing_V.V_DivergingTime}
\end{figure}

{\it Evidence for a transition, diverging time scales--}For sufficiently large $\kappa$, the non-reciprocal interactions sustain chaotic advective flows indefinitely. In contrast, for small $\kappa$, we observe that collective motion ceases abruptly after some time, and the system settles into an amorphous state with no rearrangements. In this regime, all particles drift rigidly with the same velocity and the kinetic energy in the center-of-mass frame vanishes within numerical precision: $|\mathbf{v}(t)|^2 = 1/N \sum_i |\mathbf{v}_i(t)|^2  =  0$. Because Eq.~(\ref{Eq:EqOfMotion}) is noiseless, the system cannot escape from such configuration. In the language of dynamic phase transitions, this is an absorbing state. The time $t_{\rm f}$ needed to reach an absorbing state starting from random initial conditions increases rapidly with increasing $\kappa$, until the system remains dynamically active over the entire accessible simulation time above some value $\kappa_{\rm c}$, see Fig.~\ref{fig:Freezing_V.V_DivergingTime}(a). We estimate the average freezing time $\langle t_{\rm f} \rangle$ by averaging over 20 independent initial positions, and its rapid growth towards $\kappa_{\rm c}$ is shown in Fig.~\ref{fig:Freezing_V.V_DivergingTime}(c). 

To characterize the dynamics and relevant timescales in the active phase, we consider in Fig.~\ref{fig:Freezing_V.V_DivergingTime}(b) the velocity autocorrelation $C_v(t) = \langle \mathbf{v}_i (t) \cdot \mathbf{v}_i(0)\rangle$ measured in the steady state. It is well described by a two-step exponential decay towards a nonzero plateau at long times. Mathematically, we have $C_v(t) \simeq a \exp(-t/\tau_{\rm c}) + b \exp(-t/\tau_{\rm v}) + u^2$, with $(a, b)$ quantifying the amplitude and $(\tau_{\rm c}, \tau_{\rm v})$ the timescales of the two processes. 

The initial decay at $\tau_{\rm c}$ reflects inter-particle collisions interrupting the initial ballistic motion of individual particles. This  sets a microscopic timescale in the dynamics. This is followed by a plateau, indicating persistence of motion along the streamlines shown in Fig.~\ref{fig:Model_VelocityField_DyeMix} over a timescale $\tau_{\rm v}$ that strongly depends on $\kappa$. It is remarkable that highly persistent motion emerges from many-body interactions, despite the fact that the dynamics in Eq.~(\ref{Eq:EqOfMotion}) is noiseless and introduces no persistence timescale or explicit self-propulsion. At large times, $C_v(t \to \infty) = u^2$, indicating that particles display at large timescales ballistic motion at velocity $u$ in the direction opposite to $\mathbf{u}_i$ (recall that orientations are frozen).

As $\kappa$ is reduced towards $\kappa_{\rm c}$, approaching the regime of zero kinetic activity, the particle motion slows down as a consequence of the reduced non-reciprocal drive, causing both the microscopic time $\tau_{\rm c}$ and the persistent time $\tau_{\rm v}$ to increase. We extract $\tau_{\rm c}$ and $\tau_{\rm v}$ by fitting the velocity autocorrelations to the above functional form. As shown in Fig.~\ref{fig:Freezing_V.V_DivergingTime}(c), the ratio $\tau_{\rm v}/\tau_{\rm c}$ diverges as $\kappa$ decreases towards  $\kappa_{\rm c}$, indicating the emergence of increasingly persistent velocity correlations. The divergence of two distinct timescales on opposite sides of $\kappa_{\rm c}$ provides strong evidence for the existence of an absorbing-state phase transition at a finite critical value $\kappa_{\rm c}$.

\begin{figure*}[t]
\includegraphics{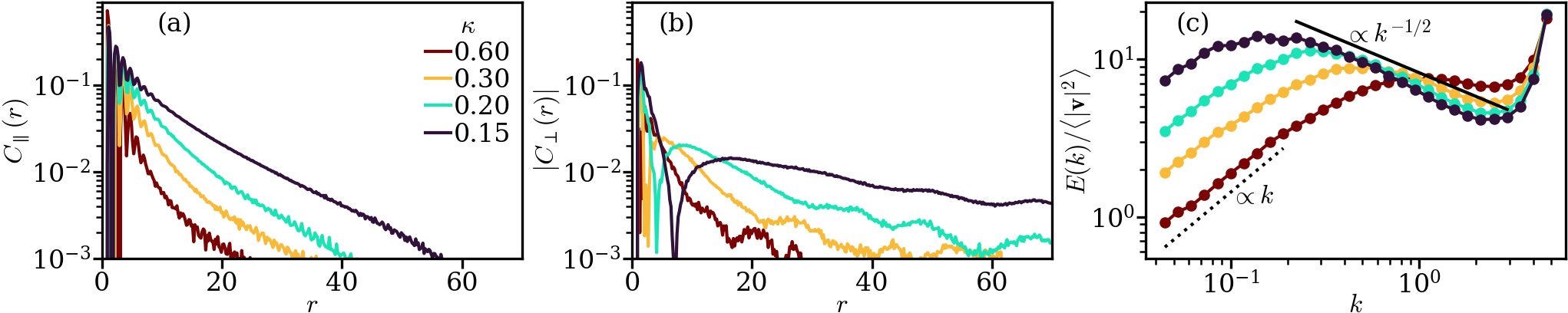}
\caption{(a) Longitudinal and (b) transverse velocity correlations, Eq.~\eqref{Eq:C(r)}, for different values of $\kappa>\kappa_{\rm c}$. (c) Kinetic energy spectrum $E(k)$, Eq.~\eqref{Eq:E(k)}, for the same parameters. All three functions demonstrate the growth of a correlation lengthscale diverging towards $\kappa_{\rm c}$, opening a regime of scale-free correlations.}
\label{fig:SpatialCorrelations}
\end{figure*}

{\it Spatial correlations and energy spectra--}We now examine the spatial correlations of the turbulent-like velocity field in the steady state regime, $\kappa >\kappa_{\rm c}$. To characterize the emergent patterns of streams and vortices, we follow Ref.~\cite{Mesoscale2024} and decompose the real-space velocity correlations into longitudinal and transverse components:
\begin{equation}
C_\lambda(r) = \frac{\langle \sum_{i,j} v_i^\lambda v_j^\lambda \delta(r_{ij}-r) \rangle}{\langle  \sum_{i,j}  \delta(r_{ij}-r)\rangle},
\label{Eq:C(r)}
\end{equation}
where $v_i^\lambda$ denotes the component of ${\bf v}_i$ either longitudinal ($\lambda = \parallel$) or transverse ($\lambda = \perp$) to ${\bf r}_{ij}$. This decomposition separates in the total correlation $C(r) = C_\parallel(r) + C_\perp(r)$ the positive correlations along streams in $C_\parallel(r)$ from negative correlations characteristic of vortices in $C_\parallel(r)$, see Figs.~\ref{fig:SpatialCorrelations}(a, b). As $\kappa \to \kappa_{\rm c}^+$, both components show a significant increase in range and amplitude, signaling the presence of large-scale collective motion over lengthscales that seem to diverge as the transition is approached. 

A Fourier space analysis is often used in active turbulent systems~\cite{alert2022active}. The kinetic energy spectrum reads
\begin{equation}
E(k) = \frac{2\pi}{L^2}k \langle|\tilde{\mathbf{v}}(\mathbf{k})|^2\rangle\,,
\label{Eq:E(k)}
\end{equation}
where $\tilde{\mathbf{v}}(\mathbf{k})=\int d^2\mathbf{r} \, \mathbf{v}\exp(-i\mathbf{k}\cdot\mathbf{r})$ is the Fourier transform of the velocity field $\mathbf{v}(\mathbf{r}) = \sum_i \mathbf{v}_i \delta(\mathbf{r}-\mathbf{r}_i)$, and $k = |\mathbf{k}|$. The data in Fig.~\ref{fig:SpatialCorrelations}(c) show that $E(k)$ exhibits a pronounced peak at wavevector $k_{\rm max}(\kappa)$, separating a linear increase $E(k) \sim k$ at small $k$ from a power-law decay above the peak, $E(k) \sim k^{-\alpha}$. The upturn at even larger $k$ reflects local correlations of the disordered packing at shor scale, $k \sim 2 \pi / \sigma$. The linear scaling at small $k$ reflects the absence of velocity correlations beyond a lengthscale $2 \pi / k_{\rm max}$. In contrast, the power-law decay $E(k)\sim k^{-\alpha}$ for $k_{\rm max} < k < 2\pi / \sigma$ indicates an emergent scale-free regime in real space, with spatial correlations decaying as $C(r) \sim r^{\alpha-1}$. The dominant wavevector $k_{\rm max}$ thus provides a direct estimate of the correlation length, which grows continuously as $\kappa$ approaches $\kappa_c$.

\begin{figure*}[t]
\includegraphics{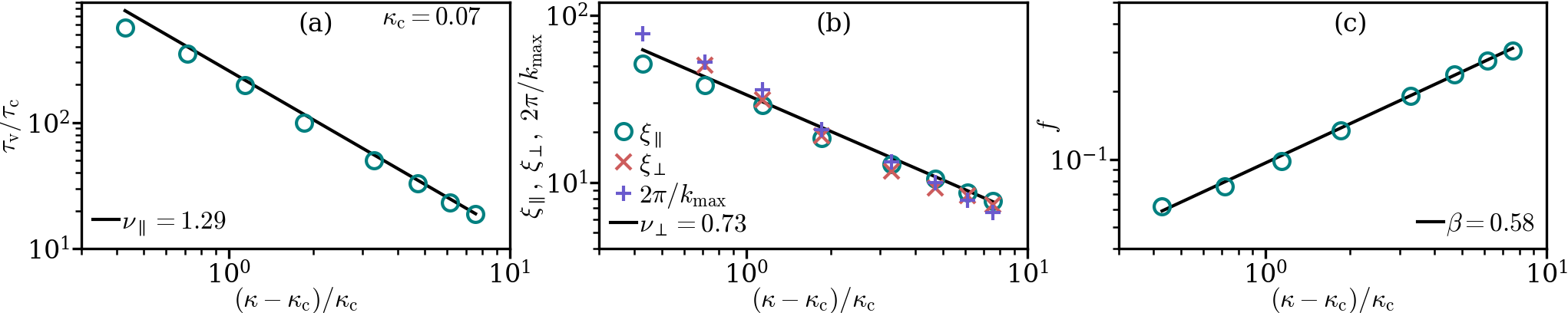}
\caption{Directed Percolation scaling laws (solid black line) for timescales (a), lengthscales (b), and the activity order parameter (c) describe the numerical data very well (symbols) using known DP values for $(\nu_\parallel, \nu_\perp, \beta)$ in two dimensions.} 
\label{fig:CriticalScaling}
\end{figure*}

{\it Directed percolation--}Absorbing phase transitions arise in a variety of systems, from epidemic spreading and catalytic reactions~\cite{Harris1974,Hinrichsen2000,ZiffBarshad1986} to driven suspensions~\cite{CortePine2008,ReichhardtReichhardt2014,TjhungBerthier2016,NessCates2020,JocteurMari2024}. In the absence of specific symmetries or conservation laws, these transitions fall into the same universality class as directed percolation (DP)~\cite{Hinrichsen2000}. This is characterized by three independent exponents~\cite{Luebeck2004} governing the divergence of timescales ($\nu_\parallel \simeq 1.29$), lengthscales ($\nu_\perp \simeq 0.73$), and the vanishing activity ($\beta \simeq 0.58$) (numbers are quoted for two dimensions).  We now demonstrate that criticality near $\kappa_{\rm c}$ in our model is consistent with DP. Starting with timescales, Fig.~\ref{fig:CriticalScaling}(a) shows the rescaled persistence time $\tau_{\rm v}/\tau_{\rm c}$, as a function of the reduced amplitude $(\kappa-\kappa_{\rm c})/\kappa_{\rm c}$, using $\kappa_{\rm c}=0.07$. The data show excellent agreement with the DP prediction for $\nu_\parallel$.  

To test the critical scaling of lengthscales, we compare three independent estimates. The longitudinal lengthscale $\xi_\parallel$ is defined as the distance at which $C_\parallel(r)$ decays to a threshold value, i.e. $C_\parallel(r=\xi_\parallel) = 10^{-2}$. Similarly, the transverse correlation length $\xi_\perp$ is obtained from the decaying negative tail of $C_\perp(r)$, i.e. $-C_\perp(r = \xi_\perp) = 10^{-2}$. Finally, $k_{\rm max}$ is determined from a scaling collapse of the energy spectrum $E(k)$, by plotting $E(k)/(k A)$ versus $k/k_{\rm max}$, where A is a constant. As shown in Fig.~\ref{fig:CriticalScaling}(b), the three quantities follow the DP prediction for lengthscales near criticality, with the exponent $\nu_\perp$. 

It is not obvious how to define `activity' in our model, compared to, say, lattice sandpile models with discrete variables~\cite{Luebeck2004}. We propose to consider the projection of the velocity ${\bf v}_i$ along the orientation ${\bf u}_i$. More precisely, we define a non-dimensional activity $f_i \equiv - {\bf u}_i \cdot {\bf v}_i / |{\bf v}_i|$. When $\kappa$ is very large, particles move mostly in the direction opposite to ${\bf u}_i$ and $f_i \to 1$. As $\kappa$ decreases, the head-tail force imbalance becomes less efficient, and motion is not along $-\mathbf{u}_i$. To sustain motion, particles must instead move more collectively along transverse directions, and $f_i$ decreases. By construction, $f_i = 0$ below $\kappa_{\rm c}$. The adimensional average activity $f \equiv \langle f_i \rangle$ should thus naturally capture the continuous approach to the absorbing phase transition. In Fig.~\ref{fig:CriticalScaling}(c), we compare the power-law decay of $f$ with the predicted critical scaling of DP $f \sim (\kappa - \kappa_c)^\beta$, and find excellent agreement. As an alternative definition of global activity, we defined the ratio $\tilde{f} = u/v$ between the averaged long-time longitudinal velocity $u$ and the short-time velocity $v$, which is again large when particles move along their orientations, and vanishes at $\kappa_c$. The numerics (not shown) is also consistent with $\tilde{f} \sim (\kappa - \kappa_c)^\beta$.

{\it Discussion--}Our study demonstrates that an athermal glassy solid driven by local non-reciprocal forces undergoes a sharp transition to a flowing fluid state at a finite value of the driving amplitude. This dynamic phase transition displays a critical behavior consistent with directed percolation. These results challenge analytic predictions from dynamical mean-field theory that $\kappa_c=0$, i.e. chaotic dynamics is predicted at any value of the driving forces~\cite{SompolinskySommers1988,CrisantiSompolinsky1987,berthier2000two}. Similar discrepancies have been noted before for global forcing~\cite{berthier2025yielding}. We suggest that theoretical progress to describe depinning~\cite{Fisher1998} and yielding~\cite{berthier2025yielding} transitions in finite dimensional materials could fruitfully be extended to the case of non-reciprocal forces. 

Our study intentionally focused on the noiseless limit of Eq.~(\ref{Eq:EqOfMotion}). Two natural sources of noise would be thermal fluctuations and a finite persistence time for the orientations ${\bf u}_i$. While a frozen glassy state will qualitatively survive, the presence of noise will transform the sharp athermal transition into a timescale-dependent crossover, akin to a dynamic glass transition. Noise triggers activated processes which induce slow aging, creep, or logarithmic time dependencies. We leave the study of these more complicated dynamics carefully for future work. 

Another line of research would consider a broader family of driving forces, to confirm that our findings are generic to a broad class of locally-driven dense particle assemblies. It would also be interesting to analyze the case of monodisperse systems which can form ordered solids. Here, driving forces should disrupt ordering and possibly melt the solid state, as we could observe in preliminary studies. The interplay between crystalline order and other types of non-reciprocal forces has been studied experimentally recently~\cite{BililignIrvine2022,GuilletBartolo2025}, and it would be interesting to compare our microscopic model with these different types of forcing that use a rotating magnetic field or hydrodynamic flows. Implementing alternative sources of non-reciprocity—such as self-propulsion rules coupled to neighbor positions~\cite{KhadakaCichos2018}—would also be interesting. Our observation that the chaotic steady state above threshold displays signature of large-scale velocity correlations and energy spectra characteristic of active turbulence suggests a possible experimental route to design artificial active systems with controlled and tunable emergent properties at large scale both in two and three spatial dimensions.  
 
\acknowledgments 
We thank R. Golestanian, D. Levis, S. Loos, and S. Ramaswamy for discussions. This research was supported in part by grant NSF PHY-2309135 to the Kavli Institute for Theoretical Physics (KITP). L.B. acknowledges the support of the French Agence Nationale de la Recherche (ANR), under grants ANR-20-CE30-0031 (project THEMA) and ANR-24-CE30-0442 (project GLASSGO).

\bibliography{refs}

\end{document}